\documentstyle[epsfig,amssymb,12pt]{article}

\begin{document}

\begin{titlepage}

\begin{center}

\mbox{\Large \bf Precise critical exponents for the basic contact process}

\vspace{4.0pc}

{\large  J Ricardo G de Mendon\c{c}a\footnote{Electronic address:
jricardo@power.ufscar.br}}

\vspace{0.5pc}

\mbox{\hspace{-1.0pc}Departamento de F\'{\i}sica, Universidade Federal
de S\~{a}o Carlos, 13565-905 S\~{a}o Carlos, SP, Brazil}

\vspace{4.0pc}

{\large Abstract}

\vspace{1.0pc}

\parbox{5.5in}
{We calculated some of the critical exponents of the directed percolation
universality class through exact numerical diagonalisations of the master
operator of the one-dimensional basic contact process. Perusal of the power 
method together with finite-size scaling allowed us to achieve a high degree
of accuracy in our estimates with relatively little computational effort.
A simple reasoning leading to the appropriate choice of the microscopic time
scale for time-dependent simulations of Markov chains within the so called 
quantum chain formulation is discussed. Our approach is applicable to any
stochastic process with a finite number of absorbing states.

\vspace{1.0pc}

{\noindent}PACS numbers: 64.60.Ht, 05.70.Jk, 02.50.Ga

\vspace{1.0pc}

{\noindent}Keywords: contact process, absorbing states, power method}

\end{center}

\end{titlepage}

\newpage

In its original formulation \cite{zpb42,zpb47}, the so-called directed
percolation (DP) conjecture stated that all continuous phase transitions
about a single absorbing state in single-component systems
with a scalar order parameter are in the DP universality class of critical
behaviour \cite{kinzel}. In this form the conjecture has been confirmed
in a host of model systems, amongst others the basic contact process 
\cite{harris,liggett}, Schl\"{o}gl's models for autocatalytic chemical 
reactions \cite{zpb42,zpb47,schlogl}, and a phenomenological classical 
field theory of high energy hadronic collision processes 
\cite{brower,reggeon,delatorre,cardysugar}. Further investigation revealed
that the DP universality class is robust enough to accommodate more general 
models, with more than a single component \cite{zgb,chopard,grinstein,fogedby}
as well as with multiple, in some cases infinitely many absorbing states 
\cite{dickman,albano,munoz}. Even some nonequilibrium growth models
without absorbing states were found to share some of the DP exponents
\cite{kertesz,alon,jrgm}. For recent reviews see \cite{marro,granada}.

The basic contact process (CP) \cite{harris,liggett} may be viewed as a model
for the spread of an epidemic amongst individuals living in a $d$-dimensional 
orchard. In this process, a healthy individual $\emptyset$ becomes infected
at a rate proportional to the number of its infected neighbours, whilst an 
infected individual $X$ becomes healthy at unit rate. Pictorially, in one 
dimension it is defined by the elementary processes 
$X \emptyset \emptyset \stackrel{\lambda}{\to} X X \emptyset$, 
$\emptyset \emptyset X \stackrel{\lambda}{\to} \emptyset X X$, 
$X \emptyset X \stackrel{2\lambda}{\to} X X X$, and 
$X \stackrel{1}{\to} \emptyset$. As $\lambda$ increases from zero,
the basic CP suffers an extinction-survival phase transition in all
dimensions, the upper critical dimension being $d^{*}=4$. There is not 
an exact evaluation of the critical points $\lambda^{*}$ for $d < d^{*}$
to date, but there are some narrow bounds: in one dimension it is known
that $1.539 < \lambda^{*} < 1.942$ \cite{bounds}. 

In this work we were concerned with the accurate determination of the critical
point and some of the critical exponents of the one-dimensional basic CP 
through exact numerical diagonalisations of its master operator. Our method 
is based on the standard matrix power method \cite{matrix} applied to a 
discrete-time version of the continuous-time Markov chain, taking advantage 
of the presence of an absorbing state. Combined with finite-size scaling 
\cite{fss} and modern extrapolation techniques \cite{bst}, the method allowed
for a high degree of numerical accuracy within quite reasonable computational
efforts. Successful attempts at the application of phenomenological
renormalisation group ideas to directed problems on the lattice dates back
at least to the work of Kinzel and Yeomans \cite{kinzel}, and have from time
to time reappeared in the literature with varied levels of sophistication
\cite{henkel}. The application of the `quantum chain formulation' of Markov
chains on the lattice \cite{adhr} to study time-dependent properties, however,
remained scarce; for a recent example, see \cite{cieplak}. In this work we 
provide a simple reasoning leading to the appropriate choice of the microscopic
time scale for simulations that should be of value to anyone interested in
similar calculations.

The starting point of our approach is the master equation on the lattice.
Let $\Lambda \subset {\Bbb Z}$ be a one-dimensional lattice of $|\Lambda|=L$
sites with periodic boundary conditions. To each site $\ell \in \Lambda$ we
attach a random variable $n_{\ell}$ taking values in a finite set $\omega =
\{0,1,\ldots,N-1\} \subset {\Bbb N}$, the state of the whole lattice being
given by ${\bf n}=(n_{1},n_{2},\ldots,n_{L}) \in \Omega = \omega^{\Lambda}$.
In the basic CP sites can only be healthy or infected, thence $N=2$. 
Given positive real numbers $\Gamma(\tilde{\bf n},{\bf n})$ denoting the
rates at which the collision ${\bf n} \to \tilde{\bf n}$ occurs, the master
equation governing the time evolution of the probability $P({\bf n},t)$ of 
realisation of the configuration ${\bf n}$ at instant $t$ reads
\begin{equation}
\label{MASTER}
\frac{{\rm d}}{{\rm d}t}P({\bf n},t) = \sum_{\tilde{\bf n}}
\big[ \Gamma({\bf n},\tilde{\bf n})P(\tilde{\bf n},t)-
      \Gamma(\tilde{\bf n},{\bf n})P({\bf n},t) \big].
\end{equation}
We now introduce vector spaces in the description of the above equation.
To do this we turn $\omega$ into ${\Bbb C}^{N}$ and write
\begin{equation}
|P(t)\rangle = \sum_{{\bf n}}P({\bf n},t)|{\bf n}\rangle
\end{equation}
for the generating vector of the probabilities 
$P({\bf n},t) = \langle {\bf n}|P(t)\rangle$, with $\{ |{\bf n}\rangle\}$
the orthonormal basis diagonal in the occupation number representation. 
We are in this way providing the space of generating functions with a 
Hilbert space structure. We then rewrite Eq.~(\ref{MASTER}) as
\begin{equation}
\frac{{\rm d}}{{\rm d}t}|P(t)\rangle = -H|P(t)\rangle
\end{equation}
where $H$ is given by
\begin{equation}
\label{OPERATOR}
H=\sum_{\tilde{\bf n}}\sum_{{\bf n}}H(\tilde{\bf n},{\bf n})
|\tilde{\bf n}\rangle \langle{\bf n}|
\end{equation}
with $H(\tilde{\bf n},{\bf n})=-\Gamma(\tilde{\bf n},{\bf n})$ and
$H({\bf n},{\bf n})=\sum_{\tilde{\bf n} \neq {\bf n}}
\Gamma(\tilde{\bf n},{\bf n})$. $H$ is but the infinitesimal generator of 
the Markov semigroup $U(t)=\exp (-tH)$ of the continuous time Markov chain 
$\{{\bf n}(t), t \geq 0 \}$ defined by the set of rates 
$\Gamma(\tilde{\bf n},{\bf n})$. The spectrum of $H$ lies in the complex 
right half-plane, and since $H$ is a real matrix, its eigenvalues are 
either real or come in complex conjugate pairs. The steady state of $H$ has
eigenvalue zero, and for finite systems it is, up to symmetry degeneracies,
unique. If we realise the algebra of operators of $\Omega$ in terms of spin
$S$ Pauli matrices \cite{adhr}, the master operator $H$ of the basic CP can
be seen to be equivalent to a spin $S=\frac{1}{2}$, three-body non-Hermitian
quantum chain.

The lowest gap $E^{(1)}-E^{(0)}=E^{(1)}$ in the spectrum of $H$ may be
used to perform a finite-size scaling analysis in the same way as one is
used to do in equilibrium problems \cite{fss}. Let us briefly review some
formul\ae. Around the critical point $\lambda \gtrsim \lambda^{*}$, the
correlation lengths of the infinite system behave like
\begin{equation}
\label{CORR}
\xi_{\|} \propto \xi_{\perp}^{z} \propto 
(\lambda -\lambda^{*})^{-\nu_{\|}} \propto 
(\lambda -\lambda^{*})^{-\nu_{\perp}z}
\end{equation}
where $\xi_{\|}$ and $\xi_{\perp}$ are the correlation lengths respectively
in the time and space directions, $\nu_{\|}$ and $\nu_{\perp}$ are the 
corresponding critical exponents, and  $z = \nu_{\|}/\nu_{\perp}$ is the 
dynamic critical exponent. Notice that in the interacting particle systems
literature it is more usual to find $z$ defined as $z=2\nu_{\perp}/\nu_{\|}$.
For finite systems of size $L$, according to the usual finite-size scaling 
assumptions \cite{fss} we expect
\begin{equation}
\label{CSI}
\xi_{\|,L}^{-1} = 
L^{-z_{L}}\Phi\left( |\lambda -\lambda^{*}_{L}|L^{1/\nu_{\perp ,L}} \right)
\end{equation}
where $z_{L}$ and $\nu_{\perp ,L}$ are the finite versions of $z$ and 
$\nu_{\perp}$ and $\Phi(x)$ is a scaling function with $\Phi(x \gg 1) \sim 
x^{\nu_{\|}}$. On general grounds one expects 
$\lim_{L \to \infty} \lambda^{*}_{L},z_{L},\nu_{\perp ,L} = \lambda^{*},
z ,\nu_{\perp}$. From Eqs.~(\ref{CORR}) and (\ref{CSI}) we obtain
\begin{equation}
\label{PCTHETA}
      \frac{\ln \left[ \xi_{\|,L}(\lambda^{*}_{L})/\xi_{\|,L'}(\lambda^{*}_{L})
               \right]}{\ln (L/L')}  =
      \frac{\ln\left[ \xi_{\|,L''}(\lambda^{*}_{L})/\xi_{\|,L}(\lambda^{*}_{L})
                \right]}{\ln(L''/L)} =
      z_{L}
\end{equation}
which through the comparison of three different system sizes $L' < L < L''$ 
furnishes simultaneously $\lambda^{*}_{L}$ and $z_{L}$. According to
Eq.~(\ref{CSI}), the exponent $\nu_{\perp ,L}$ may be calculated through 
$\nu_{\perp ,L}^{-1}=z_{L}+\frac{1}{2}(\zeta_{L',L}+\zeta_{L,L''})$, 
where
\begin{equation}
\label{ZETA}
\zeta_{M,N}=
\frac{ \ln \left[ \left( \partial \xi_{\|,N}^{-1} / \partial \lambda \right)
           \Big/  \left( \partial \xi_{\|,M}^{-1} / \partial \lambda \right) 
           \right] }
     {\ln (N/M)}
\end{equation}
with the derivatives calculated at $\lambda = \lambda^{*}_{L}$. 
Of course $\xi_{\|,L}^{-1}={\rm Re}\{ E_{L}^{(1)}\}$.

Several possibilities exist to proceed with the calculation of the gaps.
When $H$ is a symmetric matrix, Lanczos diagonalisation 
becomes the method of choice. For stochastic processes, however, $H$ is 
in general non-symmetric, and although there exist non-symmetric variations
of the Lanczos algorithm, they are either much more memory demanding or are
intrinsically unstable \cite{matrix}. Since we are interested in only
one very precise eigenvalue, we choose to work with the power method,
which requires only matrix-by-vector multiplications that can be carried 
out using high precision data types.

In order to apply the power method, we first define the matrix 
$T=1-\tau H$. This matrix should be viewed as the time evolution operator 
of a discrete-time Markov chain whose transitions take place at intervals 
$\tau$. For $T$ to be a stochastic matrix, its elements ought to satisfy
$0 \leq T(\tilde{\bf n},{\bf n}) \leq 1$ and 
$\sum_{\tilde{\bf n}}T(\tilde{\bf n},{\bf n})=1$. This last condition is
always satisfied, simply because 
$\sum_{\tilde{\bf n}}H(\tilde{\bf n},{\bf n}) = 0$. The first condition,
however, demands that $\tau^{-1} \geq \max \{ |H(\tilde{\bf n},{\bf n})|\} =
\max \{ H({\bf n},{\bf n}) \}$. Since we will calculate the eigenvalues
of $T$ by an iterative procedure, convergence will occur at a maximal
rate if we choose $\tau$ as large as possible, and we take
$\tau^{-1} = 1.01 \times \max \{ H({\bf n},{\bf n}) \}$. It is not advisable 
to take $\tau^{-1} = \max \{ H({\bf n},{\bf n}) \}$ because this will zero 
some elements in the diagonal of $T$, thus introducing cycles in an otherwise
acyclic Markov chain. The spectrum of $T$ lies in the unit circle, with the
steady state corresponding to the eigenvalue one. Notice that the spectra
of $H$ and $T$ appear in reverse order. The above choice of the microscopic
time scale $\tau$ for the discrete-time Markov chain is equivalent to the 
requirement that the probability of two transitions taking place in time 
$\tau$ is negligible, of $o(\tau)$. This makes the approximation 
$U(\tau ) = \exp (-\tau H) \simeq 1-\tau H$ to the Markov semigroup exact
in what regards the dynamics, i.e., it preserves stochasticity, and the 
time-evolved vectors $|P(t+\tau )\rangle = \exp (-\tau H)|P(t)\rangle$ and
$|P(t+\tau )\rangle = (1-\tau H)|P(t)\rangle$ coincide up to $o(\tau)$.

The construction of the matrix $T$ presented above together with the 
appropriate choices of $\tau$ with the purpose of studying time-dependent
properties of stochastic processes is well known in queueing theory, where
with some minor refinements it is called the method of uniformisation
\cite{queue}. 

In the basic CP in finite volume, the steady state is given by the unique
absorbing state $|{\bf 0}\rangle = |0,0,\ldots,0\rangle$. This {\it a priori\/}
knowledge of the steady state of the process makes it possible to calculate
the second largest eigenvalue of $T$, that corresponding to the gap of $H$,
by simply orthogonalising the iterated vector of the power method against
the steady state at every iteration. We thus end up with an implementation
of the power method that reads
\begin{equation}
|P(m+1)\rangle = \big( T-|{\bf 0}\rangle\langle{\bf 0}|\big)|P(m)\rangle
\end{equation}
with $m$ in units of $\tau$. We expect that after enough successive
applications of the above relation it reaches a fixed point 
$E^{*}|P^{*}\rangle = T|P^{*}\rangle$ where $E^{*} = 
\langle P^{*}|T|P^{*}\rangle / \langle P^{*}|P^{*}\rangle = 1-\tau E^{(1)}$
and $|P^{*}\rangle$ is a linear combination of one-particle states.
The advantage of being dealing with an absorbing state in contrast to, e.g.,
a numerically determined steady state, is that the former is usually a `pure'
state, as is our case, or a rather simple combination of states with known
coefficients, e.g., white noise, a fact that both minimises the inevitable 
numerical round-off errors and saves computer time, turning the calculations
more reliable and fast. In our calculations, we consider an eigenvalue to 
have converged if it coincides more than 64 times with its predecessors in
more than one part in $2^{112} \simeq 5.2 \times 10^{33}$. 

\begin{figure}[t]
\vspace{-0.5in}
\begin{center}
\mbox{\epsfig{figure=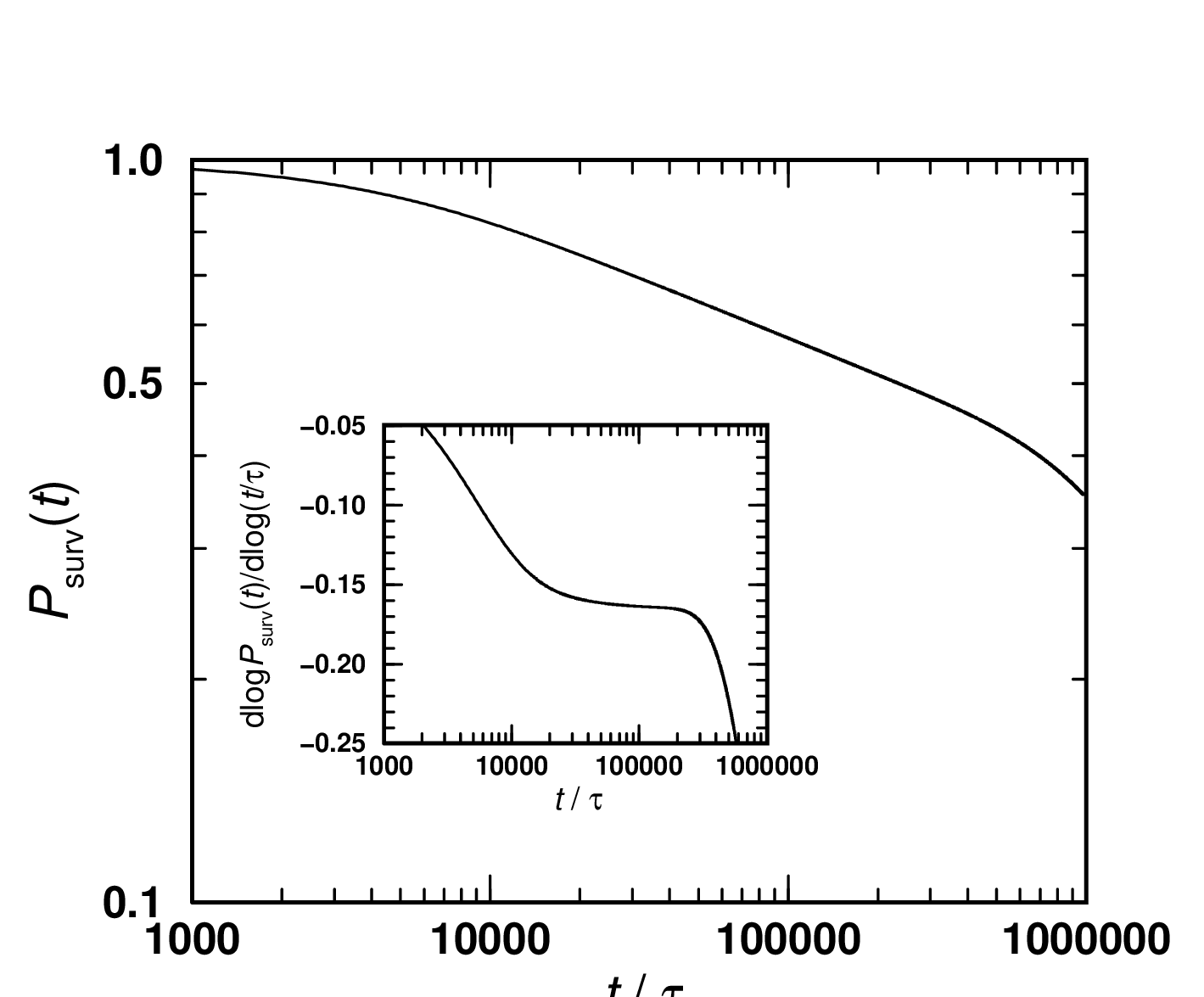,height=5.0in,width=5.0in,angle=-90}}
\end{center}
\caption{Survival probability at $\lambda = \lambda^{*}_{L}$ for a system of
$L=16$ sites. The inset shows the instantaneous value of the critical exponent
$\delta_{L}$.}
\label{FIGURE}
\end{figure}

The complete characterisation of the DP universality class requires,
besides $z$ and $\nu_{\perp}$, one more exponent to be calculated, the
other exponents following from well known hyperscaling relations 
\cite{delatorre,dickman,marro}. One calculable exponent is $\delta$, 
defined through the asymptotic behaviour of the survival probability
at the critical point $\lambda^{*}$ as
\begin{equation}
P_{\rm surv}(t) = \sum_{{\bf n} \neq {\bf 0}}P({\bf n},t) = 1-P({\bf 0},t)
\stackrel{t \to \infty}{\propto} t^{-\delta}(1+at^{-\delta'}+\ldots).
\end{equation}
A logarithmic plot of $P_{\rm surv}(t)$ versus $t$ for an infinite system
should be a straight line at large $t$, the slope of which is $\delta$. The
correction exponent $\delta'$ seems to be given by $\delta'=1$ 
\cite{marro,delta}. For finite systems, however, the spectrum always has a
gap, and the survival probability ultimately enters a regime of exponential 
decay ruled by the finite gap. In order to follow the time evolution of the 
process before it gets trapped into an absorbing state, we take a small 
$\tau$ in the definition of $T$ and successively calculate $|P(m+1)\rangle = 
T|P(m)\rangle$. We choose $\tau^{-1}=1000 \times \max\{ H({\bf n},{\bf n})\}$.
The initial state was given by $|P(0)\rangle = |1,0,\ldots ,0\rangle$, the
state with a single particle located at the origin. A plot of $P_{\rm surv}(t)$
versus $t$ for a system of $L=16$ sites is shown in Fig.\ \ref{FIGURE}. From
that figure we clearly see that after an initial transient in which the 
`highest modes' are washed out, the curve enters a regime of almost pure
algebraic decay until the gap in the spectrum manifests itself and we begin
to observe the expected late times exponential behaviour. This is more clearly
appreciated in the inset of Fig.\ \ref{FIGURE}, which shows the derivative of 
$\log P_{\rm surv}(t)$ with respect to $\log t$, that is the instantaneous
value~of~$\delta_{L}$.

Our results for $\lambda^{*}$, $z$, $\nu_{\perp}$ and $\delta$ are summarised
in Table~\ref{TABLE}. The $L=\infty$ values in this table were obtained 
through the Bulirsch-Stoer (BST) extrapolation scheme \cite{bst}, with 
$\omega_{\rm BST}$ the free parameter of the algorithm chosen so as to 
minimise the difference between the penultimate entries in the BST 
extrapolation tableaux. The derivatives in Eq.\ \ref{ZETA} were calculated
with a nine points symmetric difference formula with an $O(h^{9})$ error 
using steps of $h=10^{-9}$ \cite{singer}, whilst the values of $\delta$
were obtained through linear regression fits to $\log P_{\rm surv}(t)$ in
the region of algebraic decay of $P_{\rm surv}(t)$, typically through 100
data points with points separated by a time interval $\Delta t=100\tau$. 
The numbers in Table \ref{TABLE} confirm that the basic CP belongs to the
DP universality class of critical behaviour with a high degree of accuracy.
The uncertainties associated with the extrapolated numbers are mainly due
to finite-size effects and corrections to scaling, as well as to the
extrapolation procedure itself. Our numbers compare well with those obtained
by other means, namely, high-temperature expansions on a closely related 
`reggeon quantum spin chain' model \cite{brower}, other finite-size scaling
studies \cite{henkel}, time-dependent operator perturbation calculations
\cite{power}, and extensive series expansions and differential approximants
analyses \cite{jensen}. The best values of $z$, $\nu_{\perp}$, and $\delta$
to date are given by $z=1.580\,745(10)$, $\nu_{\perp}=1.096\,854(4)$, and 
$\delta =0.159\,464(6)$ \cite{jensen}. Our estimate of $\lambda^{*}$ is as
precise as those available in the current literature on the basic CP
\cite{bounds,power}. Although our determination of $\delta$ is not as precise
as that of $z$ and $\nu_{\perp}$, it nevertheless is precise enough to
discriminate amongst the universality classes that are likely to arise in
systems with absorbing states \cite{marro}. An alternative to estimate 
$\delta$ is to determine each $\delta_{L}$ as the value of the derivative
of the corresponding $\log P_{\rm surv}(t)$ with respect to $\log t$ at its
inflexion point, see Fig.\ \ref{FIGURE}. This provides a more `local'
determination of $\delta_{L}$ than that obtained by fitting a straight line
to $\log P_{\rm surv}(t) \times \log t$ over tens or hundreds of points, and
may improve the final result once the uncertainties are correctly assessed.
We intend to pursue this alternative in a future, more thorough study.
\begin{table}[tb]
\centering
\caption{Finite-size data and extrapolated values for the the critical point
$\lambda^{*}$ and the exponents $z = \nu_{\|}/\nu_{\perp}$, $\nu_{\perp}$ and
$\delta$ of the one-dimensional basic CP. The numbers between parentheses 
represent the estimated errors in the last digit of the data, whilst those 
data without an associated error are numerically precise to the figures 
shown.}
\label{TABLE}
\begin{tabular}{ccccc}                  \hline
  $L',L,L''$
& $\lambda^{*}_{L}$ 
& $z_{L}$
& $\nu_{\perp,L}$
& $\delta_{L}$                        \\ \hline
7,8,9    & 1.629\,092\,086\,131
         & 1.495\,084\,128\,194
         & 0.963\,208\,351\,697  
         & 0.1657(2)                  \\
8,9,10   & 1.632\,522\,345\,029
         & 1.502\,980\,235\,818
         & 0.977\,844\,866\,308 
         & 0.1656(2)                   \\
9,10,11  & 1.635\,178\,201\,359
         & 1.509\,743\,775\,238
         & 0.989\,427\,140\,315
         & 0.1654(2)                   \\
10,11,12 & 1.637\,262\,542\,035
         & 1.515\,558\,577\,190
         & 0.998\,833\,401\,056
         & 0.1653(2)                   \\
11,12,13 & 1.638\,921\,714\,266
         & 1.520\,588\,023\,063
         & 1.006\,632\,974\,979
         & 0.1651(2)                   \\
12,13,14 & 1.640\,260\,494\,445
         & 1.524\,967\,656\,357
         & 1.013\,211\,430\,501  
         & 0.1649(2)                   \\
13,14,15 & 1.641\,354\,409\,414
         & 1.528\,807\,324\,384
         & 1.018\,839\,364\,776
         & 0.1647(1)                   \\
14,15,16 & 1.642\,258\,557\,889
         & 1.532\,195\,497\,121
         & 1.023\,712\,402\,567
         & 0.1645(1)                   \\  
15,16,17 & 1.643\,013\,687\,274
         & 1.535\,203\,516\,105
         & 1.027\,975\,558\,429
         & 0.1644(1)                   \\
16,17,18 & 1.643\,650\,350\,303
         & 1.537\,889\,180\,610
         & 1.031\,738\,664\,206
         & 0.1643(1)                   \\
17,18,19 & 1.644\,191\,762\,995
         & 1.540\,299\,611\,765
         & 1.035\,086\,466\,706
         & 0.1641(1)                   \\
18,19,20 & 1.644\,655\,789\,991
         & 1.542\,473\,490\,560
         & 1.038\,085\,430\,711
         & 0.1640(1)                   \\  \hline
$L=\infty$   & 1.648\,96(2)
             & 1.580\,77(2)
             & 1.096\,81(2)
             & 0.162(2)                \\
{[}$\omega_{\rm BST}${]} & [1.071] & [1.171] & [0.895] & [2.701] \\ \hline
\end{tabular}
\end{table}

A remark about symmetries. In the calculation of the gaps of $H$, one
can take advantage of any symmetries of the process to reduce $H$ to 
block-diagonal form. The numerical diagonalisation is then performed in the
sector of lowest gap with a considerable economy of computation. Internal 
symmetries, like $U(1)$ and $Z(N)$ symmetries, usually separate the dynamics
into sectors corresponding to the closed classes of the stochastic process,
with each block a stochastic transition matrix governing the dynamics 
within the given sector. Geometric symmetries, however, like translation
and reflexion symmetries, generally lead to block operators that are not 
stochastic due to the occurrence of `artificial' combinations of states. 
If one decides to make use of geometric symmetries, care should then be 
exercised in properly weighting the basis vectors in order to interpret 
the resulting symmetry-invariant $|P(t)\rangle$ as a vector of probabilities.
In this work we explored the translational invariance of the basic CP on a 
periodic lattice in order to achieve a reduction of order $1/L$ on the size
of the matrices we needed to diagonalise.

In summary, we have successfully applied the power method to the 
one-dimensional basic CP, and obtained accurate values for the critical 
point $\lambda^{*}$ and the critical exponents $z$ and $\nu_{\perp}$,
together with a good estimate of the critical exponent $\delta$. The method
is fast, yields accurate estimates for the critical point and some of the 
exponents, and is easily coded. Given that it took less than 500 h of 
CPU time (running at 300 MHz) to complete Table \ref{TABLE}, and that in the
largest cases it took less than 21 Mb of memory to conduct the calculations,
the method seems to be very competitive. Extension to processes with more than
one absorbing state as well as in more than one dimension is immediate. It 
would be of interest to refine the calculations of $\delta$ as well as to 
try to calculate other exponents by the same methods. In particular, it would
be very interesting to establish relationships between dynamical exponents 
like $\delta$ and the spectrum of $H$. We are at the moment pursuing these 
objectives, and intend to release our results soon.

\bigskip

The author would like to acknowledge Professor Francisco C Alcaraz and
Professor Malte Henkel for useful comments on the manuscript. This work
was supported by the Funda\c{c}\~{a}o de Amparo \`{a} Pesquisa do Estado
de S\~{a}o Paulo (FAPESP), Brazil.

\end{document}